\documentclass[aps,prl,showpacs,superscriptaddress,amssymb,twocolumn]{revtex4-1}
\usepackage{graphicx}
\usepackage{appendix} 
\usepackage{bm,amsmath}
\usepackage{dcolumn}

\usepackage{xcolor}

\usepackage[citecolor=blue]{hyperref}

\begin{document}

\rightline{\tt }

\vspace{0.2in}

\title{Quantum Simulation of Operator Spreading in the Chaotic Ising Model}

\author{Michael R. Geller}
\thanks{mgeller@uga.edu}
\affiliation{Center for Simulational Physics, University of Georgia, Athens, Georgia 30602, USA}

\author{Andrew Arrasmith} 
\affiliation{Theoretical Division, Los Alamos National Laboratory, Los Alamos, NM, USA.}

\author{Zo\"{e} Holmes} 
\affiliation{Information Sciences, Los Alamos National Laboratory, Los Alamos, NM, USA.}

\author{Bin Yan} 
\affiliation{Center for Nonlinear Studies, Los Alamos National Laboratory, Los Alamos, NM, USA.}
\affiliation{Theoretical Division, Los Alamos National Laboratory, Los Alamos, NM, USA.}

\author{Patrick J. Coles}
\affiliation{Theoretical Division, Los Alamos National Laboratory, Los Alamos, NM, USA.}

\author{Andrew Sornborger} 
\thanks{sornborg@lanl.gov}
\affiliation{Information Sciences, Los Alamos National Laboratory, Los Alamos, NM, USA.}

\date{July 8, 2021}

\begin{abstract}
There is great interest in using near-term quantum computers to simulate and study foundational problems in quantum mechanics and quantum information science, such as the scrambling measured by an out-of-time-ordered correlator (OTOC). Here we use an IBM Q processor, quantum error mitigation, and {\it weaved} Trotter simulation to study high-resolution operator spreading in a 4-spin Ising model as a function of space, time, and integrability. Reaching 4 spins while retaining high circuit fidelity is made possible by the use of a physically motivated fixed-node variant of the OTOC, allowing scrambling to be estimated without overhead. We find clear signatures of ballistic operator spreading in a chaotic regime, as well as operator localization in an integrable regime. The techniques developed and demonstrated here open up the possibility of using cloud-based quantum computers to study and visualize scrambling phenomena, as well as quantum information dynamics more generally. 
\end{abstract}

\maketitle

A key concept in modern quantum physics is the noncommutativity of operators corresponding to physically separated local observables $W$ and $V$ caused by scrambling. Scrambling is a spreading of quantum information over many degrees of freedom, generated by chaotic unitary evolution \cite{swingle2018unscrambling}. The resulting growth of an operator $W(t) = e^{iHt} W e^{-iHt} $ in time can be diagnosed by the nonvanishing of the ``commutator'' \cite{hayden2007black,KitaevKITP15,shenker2014black,maldacena2016bound,sachdev1993gapless,sachdev2015bekenstein,nandkishore2015many,abanin2019colloquium} 
\begin{eqnarray}
&&{\rm tr} (\rho \big|   [W(t),V(0)] \big| ^2) = {\rm tr}[ \rho W(t)V(0)V(0)W(t)] \nonumber \\ 
&&+ \ {\rm tr}[\rho V(0)W(t)W(t)V(0)]- 2 \, {\rm Re} \, F(t)
\label{squared commutator}
\end{eqnarray}
in some state $\rho$ (often a thermal state). Here
\begin{eqnarray}
F(t) =  {\rm tr} [ \rho W(t)V(0)W(t)V(0) ]
\label{def WVWV OTOC}
\end{eqnarray}
is the out-of-time-ordered correlator (OTOC). The theoretical study of scrambling and OTOCs has enhanced our understanding of entanglement in condensed matter, quantum field theory, and quantum gravity \cite{larkin1969quasiclassical,garcia2017digital,halpern2018quasiprobability,swingle2018resilience,zhang2019information,yoshida2019disentangling,babbush2019quantum,alonso2019out,vermersch2019probing,daug2019detection,Yan2020information,belyansky2020minimal,Yan2020recovery,touil2021information,lin2018out,xu2019locality,fortes2019gauging,fortes2020signatures,holmes2021barren}.

Fast scrambling appears in a variety of systems, including black holes \cite{hayden2007black,KitaevKITP15,shenker2014black,maldacena2016bound} and  strange metals \cite{sachdev1993gapless,sachdev2015bekenstein}, while slow scrambling indicates a breakdown of ergodicity and thermalization \cite{nandkishore2015many,abanin2019colloquium}. However, the dynamics of information away from these two extremes, such as operator spreading in ordinary quantum matter, is less well understood \cite{lin2018out,xu2019locality}. Open problems also include predicting the scrambling generated by a given model Hamiltonian (without simulating it) \cite{xu2019locality}, the nature of scrambling in models without particle-like classical limits \cite{lin2018out,xu2019locality,fortes2019gauging,fortes2020signatures}, and determining when a fast scrambler has a dual description as a black hole  \cite{swingle2018unscrambling}. Additional questions that can be addressed with the techniques developed here include the dependence of scrambling on 
$\rho$, the dependence on $W$ and $V,$ and the difference between thermal scrambling in gapped and gapless phases, such as integer versus half-integer-spin 1d antiferromagnets (with or without the Haldane gap \cite{haldane1983nonlinear}).

Direct experimental measurement of (\ref{squared commutator}) is challenging because it requires reversing the direction of time (changing the sign of the Hamiltonian) during the experiment. But it can be simulated on a classical or quantum computer.  While classical simulation is limited to small or weakly correlated models, fault-tolerant quantum computers promise to make large-scale scrambling simulations practical. This should provide a valuable tool for quantum information science by complimenting the study of solvable models \cite{hayden2007black,KitaevKITP15,shenker2014black,maldacena2016bound,sachdev1993gapless,sachdev2015bekenstein,nandkishore2015many,abanin2019colloquium}. However, near-term quantum simulations are restricted to a small number of qubits and short circuits. Online users may also face additional restrictions that limit the number of distinct quantum circuits that can be measured, and thus the overall complexity of an experiment and the resulting data quality.

The quantum simulation of scrambling with near-term processors is especially challenging because, at each time~$t$, four Hamiltonian simulations of length~$t$ are implemented to calculate $F$, two each moving forward and backward in time. In addition, $F$ is complex so it has to be measured interferometrically \cite{swingle2016measuring,zhu2016measurement,yao2016interferometric}, requiring extra qubits and gates, or via weak measurement \cite{halpern2017jarzynski}, which is not widely available. It is also possible to measure scrambling via correlations between randomized measurements \cite{vermersch2019probing}. But achieving an accurate simulation over long times with high time resolution is difficult with standard Trotter simulation~\cite{cirstoiu2020variational,commeau2020variational,gibbs2021long,geller2021experimental, bharti2020iterative, lau2021quantum,haug2020generalized, barison2021efficient, trout2018simulating,endo2020variational,yao2020adaptive, benedetti2020hardware}. We introduce techniques to address these limitations and help enable the study and visualization of quantum information dynamics with cloud-based quantum computers.

The first quantum simulations of an OTOC were made by Li {\it et al.}  \cite{li2017measuring} on a nuclear magnetic resonance simulator and by G\"arttner  {\it et al.} \cite{garttner2017measuring} on a long-range Ising spin simulator. Since then, the experimental study of scrambling has made impressive progress 
\cite{meier2019exploring,nie2019detecting,landsman2019verified,blok2021quantum,chen2020detecting,niknam2020sensitivity,mi2021information,braumuller2021probing,chen2021observation,joshi2020quantum,zhu2021observation}, including recent striking demonstrations of teleportation-based OTOC measurement \cite{landsman2019verified,blok2021quantum}, which distinguishes OTOC decay due to unitary scrambling from decoherence, and Google's measurement of OTOC fluctuations on random circuits containing up to 53 qubits \cite{mi2021information}, which distinguishes operator entanglement from spreading. Google also measured operator spreading in a 2d array of qubits \cite{mi2021information}.

\begin{figure}
\includegraphics[width=9cm]{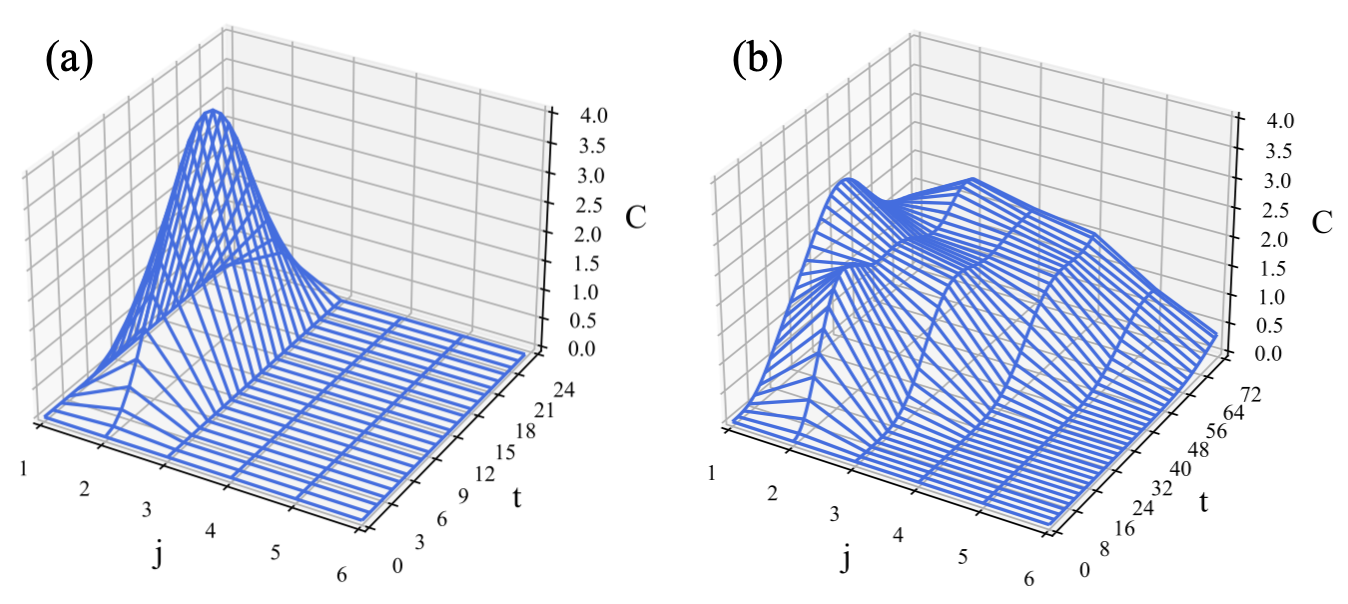} 
\caption{Commutator $C_{1j}(t)$ versus qubit position $j$ and time $t$. (a) Integrable regime, where no spreading occurs. The OTOC is calculated at times $ t \in \{ 0, \tau, 2 \tau , \cdots, 24 \tau \}$, with resolution $\tau \! = \!  0.06$, and time is plotted in units of  $\tau$.  (b) Chaotic regime, calculated with $\tau \! = \! 0.03$, showing ballistic operator spreading. Parameters for both regimes are given in Table \ref{parameter table}.}
\label{fig n6}
\end{figure} 

In this work we use quantum simulation techniques  to study spreading of the Pauli operator $X(t) \! =  \! e^{iHt} X e^{-iHt}$ in the Ising chain
\begin{eqnarray}
H &=&  H^0 + B_x  \sum_{i=1}^n  X_i  ,  
\label{defH}  \\
H^0 &=&  J  \sum_{i=1}^{n-1} Z_{i} Z_{i+1} + B_z  \sum_{i=1}^n  Z_i ,
\label{defH0}
\end{eqnarray}
with $n$ spins. The model is separated into a classical Ising chain $H^0$ plus a noncommuting transverse field. The model (\ref{defH})  permits efficient Hamiltonian simulation via Trotterization. $H^0$ is also exactly solvable, a property used below. We measure and plot the commutator
\begin{eqnarray}
C_{ij}(t) = {\rm tr}(\rho \big| [X_{i}(t), X_{j}(0)] \big|^2 ) = 2 -  2 \, {\rm Re} \, F_{ij}(t),
\label{def C}
\end{eqnarray}
in the state $\rho = |0000\rangle \langle 0000|$, with $X(t)$ initially localized at $i\!=\!1$. In particular, we study $C_{1j}$ as a function of qubit position $j \in \{1,2, \dots, N\}$ and time $t$. The quantum simulations are implemented on the IBM Q processor 
ibmq\_sydney using qubits $\{ Q_1, Q_2, Q_3, Q_5 \}$ \cite{SI}. 
The measured OTOC is
\begin{eqnarray}
F_{1j}(t) =  \langle 0000| U^\dagger X_1 U X_j  U^\dagger X_1 U X_j  |0000\rangle ,
\label{def OTOC}
\end{eqnarray}
where $U$ is a 4-qubit circuit simulating $e^{-i H t}$.  The commutator (\ref{def C}) was chosen because it exhibits a particularly smooth, easily visualized  dynamics, using an easy-to-prepare state. Operator spreading diagnosed by alternative commutators are compared in \cite{SI}.

In the experiments we always use one of the two parameter sets given in Table \ref{parameter table}, both of which simulate ferromagnetic spin chains. One set corresponds to an integrable regime of the dynamics; the other generates quantum chaos, which causes unitary OTOC decay. The models are chosen to display smooth charge spreading dynamics on the timescales of interest. The measurements in the integrable regime are relevant for recent theoretical work investigating the role of integrability on scrambling \cite{lin2018out,fortes2019gauging}, as well as serving as a scrambling-free experimental control. The commutator (\ref{def C}) is a real number $0 \le C_{1j}(t) \le 4$, and can be represented by a surface in spacetime $j \! \times \! t$. Figure \ref{fig n6} shows two such surfaces, obtained by classical simulation, for $n \! = \! 6$. These simulations assume perfect Hamiltonian simulation $U = e^{-iHt}$ with no Trotter error. Operator localization in Fig.~\ref{fig n6}a and spreading in Fig.~\ref{fig n6}b are evident. These high-resolution simulations allow one to investigate operator spreading dynamics in great detail.

\begin{figure}
\includegraphics[width=9cm]{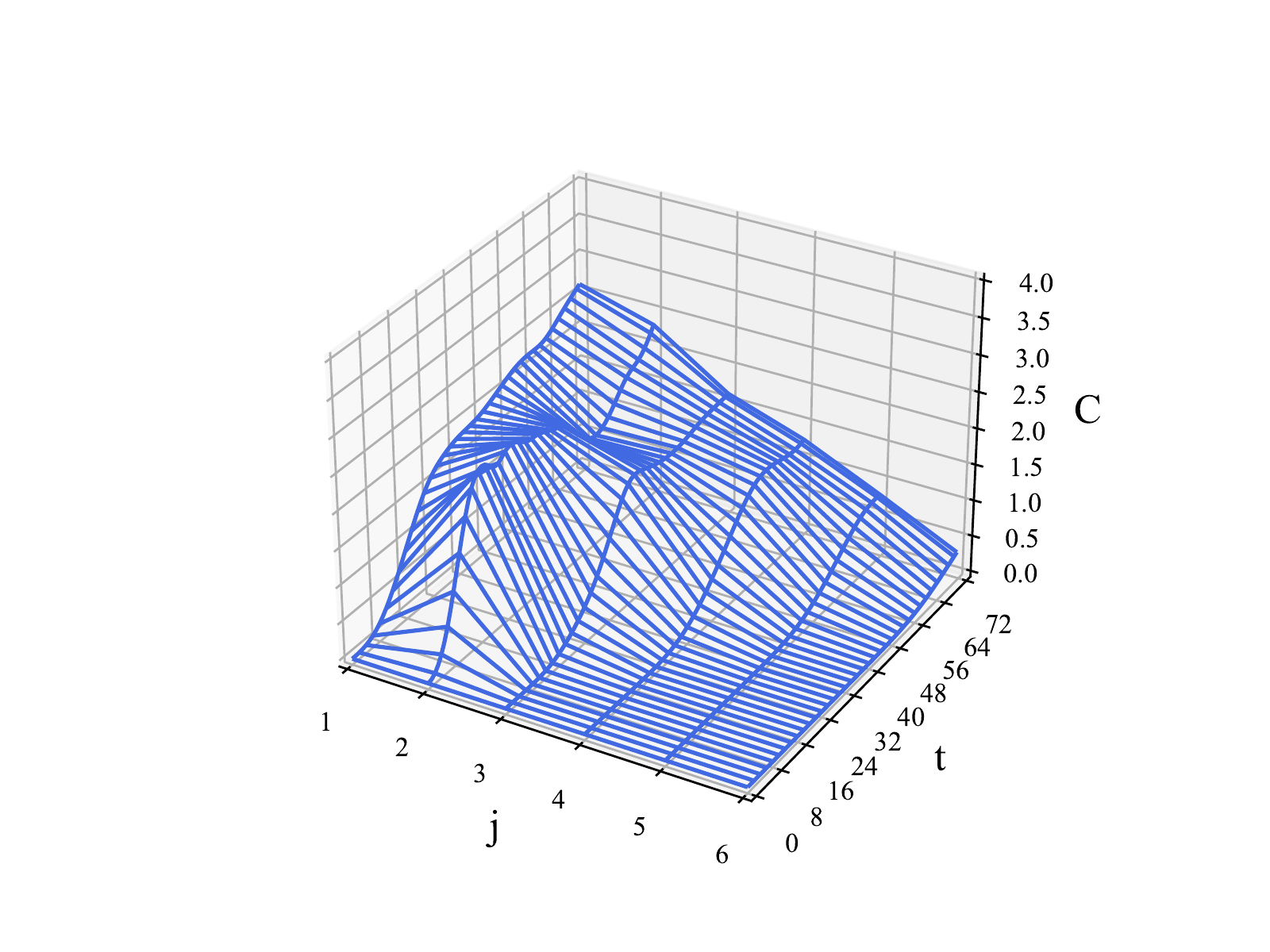} 
\caption{Chaotic $C_{1j}(t)$ commutator of Fig.~\ref{fig n6}b recalculated with the fixed-node OTOC.}
\label{fig fixednode}
\end{figure} 

\begin{table}[htb]
\centering
\caption{Ising model parameters.}
\begin{tabular}{|c|c|c|c|}
\hline
 & $J$ & $B_x$ & $B_z$ \\
\hline 
Integrable regime& -1 & 0 & 1 \\
\hline 
Chaotic regime& -1 & 0.7 & 1.5 \\
\hline 
\end{tabular}
\label{parameter table}
\end{table}

{\it Fixed-node OTOC.} A standard approach to measuring an OTOC on $n$ qubits is to add a qubit that can control the $W$ and $V$ gates \cite{swingle2016measuring,zhu2016measurement}. Here we introduce an alternative approach, which is approximate but allows one to reach larger problem sizes. Writing the OTOC in polar form as $F_{ij}  =  e^{i \arg F_{ij}} \, |F_{ij}|$, we note that the phase becomes irrelevant in the scrambling regime, because  $|F_{ij}| \approx 0$ there. $|F_{ij}|$ can be measured directly with no qubit overhead \cite{swingle2016measuring}, at least on simple states. Therefore we introduce an approximation for $\arg F_{ij}$ that is exact in the integrable regime, namely $\arg F_{ij}^0$, where $F_{ij}^0$ is the OTOC calculated with the classical Hamiltonian $H^0.$ For the Ising model (\ref{defH}), 
\begin{eqnarray}
F^0_{1j}(t) =
\begin{cases}
e^{4i(J+B_z)t} & {\rm if} \  j=1, \\
e^{4iJt} & {\rm if} \   j= 2, \\
1 & {\rm if} \   j >  2.  \\
\end{cases}
\label{F0 result}
\end{eqnarray}
This results in a {\it fixed-node} variant of the OTOC,
\begin{eqnarray}
F_{ij}  =  e^{i \arg F_{ij}^0} \, |F_{ij}| .
\end{eqnarray}
One can think of the fixed-node OTOC as an approximation to (\ref{def WVWV OTOC}), or as an independent quantity that also diagnoses scrambling.

Using the fixed-node OTOC, the commutator takes the form
\begin{eqnarray}
C_{ij} = 2 - 2  |F_{ij}| \cos(\arg F_{ij}^0).
\end{eqnarray}
The chaotic surface of Fig.~\ref{fig n6}b, recalculated with the fixed-node OTOC, is shown in Fig.~\ref{fig fixednode}.

By construction, the fixed-node OTOC is close to the exact OTOC in the late time regime, where $F_{ij} \approx 0 \ (C_{ij} \approx 2)$. However, it also satisfies an important {\it causality} constraint that extends its accuracy to the regime of $F_{ij} \approx 1 \ (C_{ij} \approx 0)$ as well, namely, the early scrambling regime. Causality requires that the exact OTOC satisfies $F_{ij}=1$ outside the {\it lightcone}, i.e. when $|{\bf r}_i-{\bf r}_j| > vt$, with ${\bf r}_{i,j}$ the local operator positions and $v$ the butterfly velocity. Therefore 
${\rm arg}(F_{ij}) \! = \! 0$ there. Because $F_{ij}^0$ also satisfies this constraint,  $C_{ij}$ computed from the fixed-node OTOC is also accurate outside the lightcone, as is evident in Fig.~\ref{fig fixednode}.
Thus, while there are small differences in the peak structure in $C_{1j}(t)$, mainly at $j \! = \! 1$, the overall spreading dynamics is accurately captured. The fixed-node commutator in the integrable regime is identical to the surface of Fig.~\ref{fig n6}a.

\begin{figure}
\includegraphics[width=8.5cm]{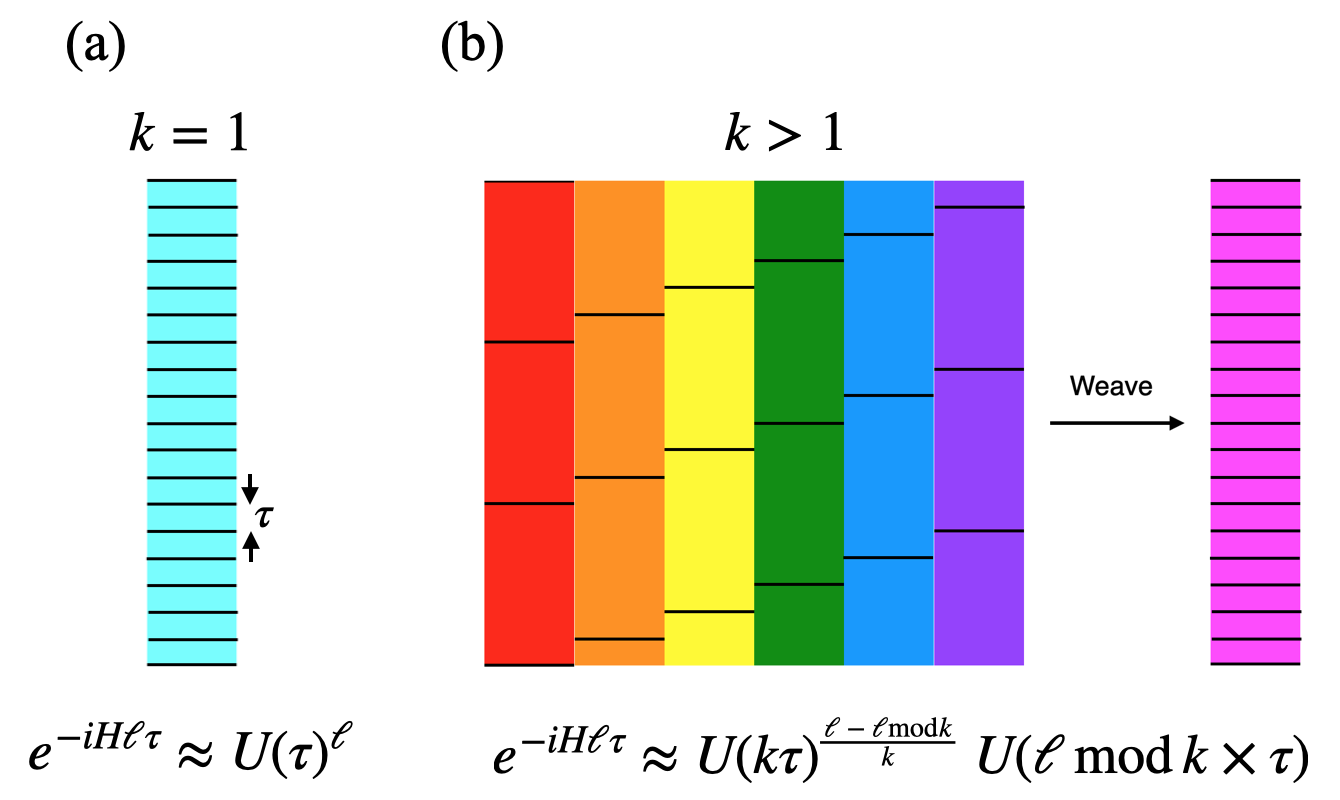} 
\caption{(Color online) Weaved time evolution. (a) Standard Trotter method, where a step $U(\tau)$ is repeatedly applied. (b) Weaving implements a fine-grained time evolution, with time resolution $\tau$, without incurring the gate errors of (a). Here the red column represents the sequence $U(k\tau) \cdots U(k\tau) $,  orange represents  $U(k\tau) \cdots  U(k\tau) U(\tau) $, yellow represents  $U(k\tau) \cdots  U(k\tau) U(2\tau) $ and purple represents  $U(k\tau) \cdots U(k\tau) \, U( (k \! - \! 1) \tau).$} 
\label{fig trotterweave}
\end{figure} 

{\it Trotter weave.} To measure $C_{ij}(t)$ with the high time resolution used in Fig.~\ref{fig n6}, one might construct a Trotter approximation $U(\tau) \approx e^{-i H \tau}$ for evolution by a short time $\tau$. Then, to simulate later times $t \! = \! \ell\tau$, the operator $U(\tau)$ is applied $\ell$ times:
\begin{eqnarray}
e^{-i H \ell \tau} \approx U(\tau)^\ell \! .
\label{k=1 weave}
\end{eqnarray}
The value of $\tau$ determines the time resolution. However, the circuit depth resulting from this standard Trotterization, based on the repeated application of a single step $U(\tau)$, is $t/\tau$.  Thus, the performance of (\ref{k=1 weave}) quickly degrades with $t$ due to gate errors and decoherence, limiting the simulation to short times.  We address this by an extension of Trotter simulation based on a collection of elementary evolution operators.  

\begin{figure}
\includegraphics[width=6.0cm]{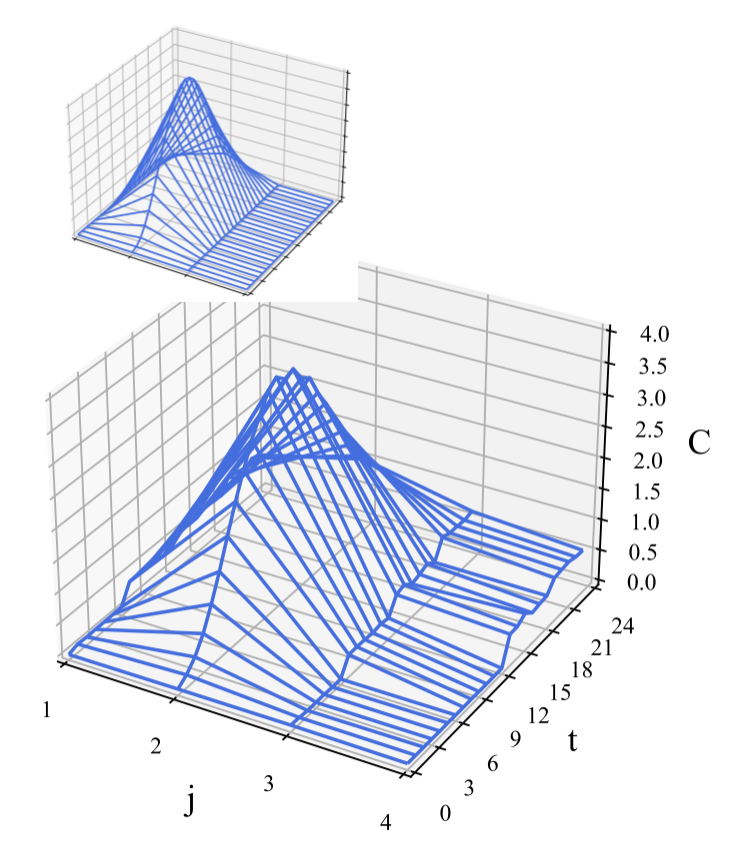} 
\caption{(Color online) Measured operator spreading surface in the integral regime, using the fixed-node OTOC. The main figure uses a 6-weave with $\tau = 0.06$, and includes both CNOT and measurement error mitigation.  
$C_{1j}$ is defined in (\ref{def C}), with $j \in \{ 1,2,3,4\} $ the coordinate of the probe qubit. The OTOC is measured at times $ t \in \{ 0, \tau, 2 \tau , 3 \tau , \cdots, 24 \tau \}$, and time is plotted in units of  $\tau$. The inset shows the classically computed noise-free surface.}
\label{fig integrable k6 corrected}
\end{figure} 

\begin{figure}
\includegraphics[width=6.0cm]{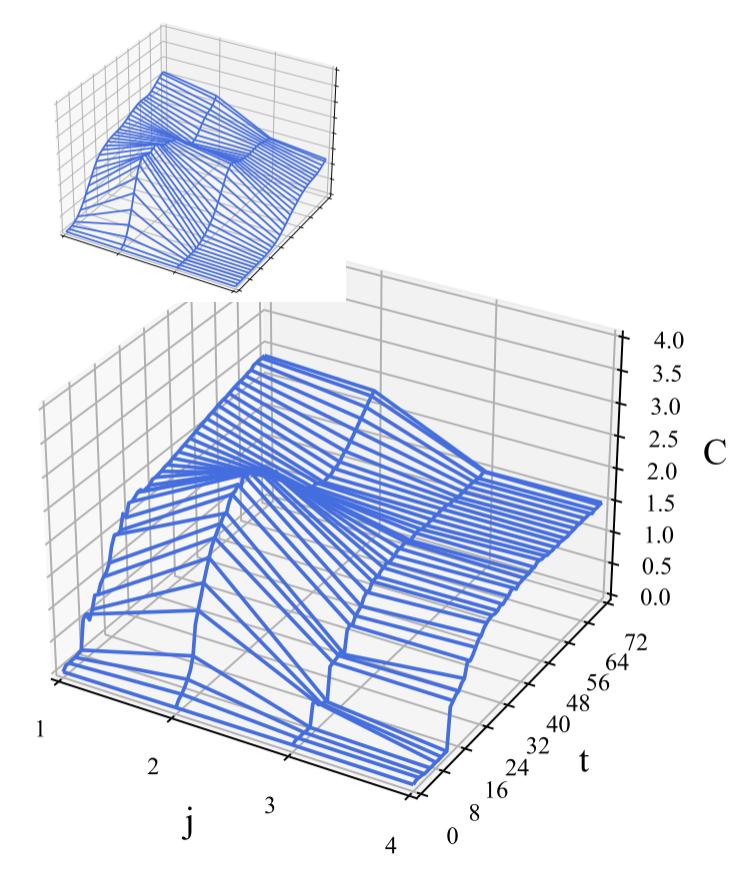} 
\caption{(Color online) Operator spreading in the chaotic regime using the fixed-node OTOC.  We use a 6-weave with $\tau = 0.03$, and also include error mitigation. The OTOC is measured at times $ t \in \{ 0, \tau, 2 \tau ,  \cdots, 72 \tau \}$. The inset shows the noise-free surface.}
\label{fig chaotic k6 corrected}
\end{figure} 

A $k$-{\it weave} is a set of $k$ unitaries
\begin{eqnarray}
\bigg\lbrace U(\tau), U(2\tau), \cdots , U(k \tau) \bigg\rbrace,
\label{def k weave}
\end{eqnarray}
where $\tau$ is a fixed short evolution time and each $U(dt)$ is a Trotterized evolution operator, propagating a state for a short time $dt$. The integer $k \in \{ 1,2,3, \cdots \}$ is called the weave modulus. The unitaries in (\ref{def k weave}) have different roles: $U(k \tau)$ is the {\it cell} operator and the remaining $k-1$ unitaries are called {\it shift} operators. Repeated applications of the cell propagates a state from $t=0$ to a time $t = (\ell - \ell \, {\rm mod} \, k) \! \times \!  \tau$, but with coarse time resolution $k \tau$. The coarse evolution is then followed (or preceeded) by a single shift operator propagating for a short interval with fine-grained time precision $\tau$. Instead of (\ref{k=1 weave}), weaving implements a time evolution according to
\begin{eqnarray}
e^{-i H \ell \tau} \approx U(k \tau)^{\frac{\ell - \ell {\rm mod} k}{k}}  U(\ell \, {\rm mod} \, k \times  \tau) ,
\end{eqnarray}
where the cell operator is applied $(\ell - \ell \, {\rm mod} \, k)/k$ times after a single shift operator. This is illustrated in Fig.~\ref{fig trotterweave}. When $\ell \gg k$, this leads to a $k$-fold reduction in circuit depth, and to higher circuit fidelity, at the expense of increased of Trotter error. Weaving can be applied to Hamiltonian simulation techniques beyond Trotterization as well.

For our first set of measurements we use the weave operators $\{ U(\ell \tau) \}_{\ell = 1}^k$, where
\begin{eqnarray}
U(\ell \tau) &=&  \prod_{i=1}^{n} \! R_{\rm x}( B_x \ell \tau )_i
 \times \prod_{i=1}^{n-1}  \! R_{\rm zz}( 2J \ell \tau  )_{i,i+1}  
 \ \nonumber \\
 &\times&
 \prod_{i=1}^{n} \! P_{\rm z}( 2 B_z \ell \tau)_i
 \times
\prod_{i=1}^{n} \! R_{\rm x}(B_x \ell \tau)_i , 
\label{trotter step}
\end{eqnarray}
resulting from a Trotterization of the model (\ref{defH}). Here $R_x(\theta)_i \! = \! e^{-i \theta X_i /2}$ is a rotation on qubit $i$, 
\begin{eqnarray}
R_{zz}(\theta)_{ij} = e^{-i\frac{\theta}{2}  Z_{i} \otimes Z_{j}}
= {\rm CNOT}_{ij} \ P_z(\theta)_i \  {\rm CNOT}_{ij} \ \ \ \ \
\label{def zz rotation}
\end{eqnarray}
is a $ZZ$ rotation, $P_z(\varphi) \! = \!  {\rm diag}(1, e^{i \varphi})$ is a phase gate, and ${\rm CNOT}_{ij}$ is an $X$ on qubit $j$ controlled by $i$. For general angles $\theta$, $R_{zz}(\theta)_{ij}$ requires two CNOTs per graph edge, or $2(n-1)$ CNOTs per Trotter step on a chain of length $n$. The quantum simulation results reported here use (up to) two weave operators per Hamiltonian simulation, requiring $4(n-1)$ CNOTs per evolution. This leads to a total of $16(n-1)$ CNOTs per OTOC simulation (48 CNOTs for $n=4$).

{\it Results.}  All of the quantum simulation results in this study were obtained with the IBM Q processor ibmq\_sydney on the 4-qubit chain $\{ Q_1, Q_2, Q_3, Q_5 \}$ \cite{SI}. Quantum simulations using Trotter weave modulus $k=6$ are shown in Figs.~\ref{fig integrable k6 corrected} and \ref{fig chaotic k6 corrected}. Figure~\ref{fig integrable k6 corrected} shows the measured spreading surface $C_{1j}(t)$ in an integral regime of the spin chain. Figure \ref{fig chaotic k6 corrected} is the same, but calculated in a chaotic regime of the model. The parameter values used are shown in Table \ref{parameter table}. The data in both regimes are corrected for measurement errors using transition matrix error mitigation \cite{maciejewski2020mitigation,nachman2020unfolding,hamilton2020scalable,bravyi2021mitigating,geller2021toward}, and for incoherent CNOT errors using zero-noise extrapolation \cite{temme2017error,dumitrescu2018cloud}. The raw data from this study is provided in \cite{SI}.

The values of $\tau$ (time resolution) and $k$ (weave modulus) are chosen to balance gate errors and Trotter errors over the relevant time scales of the simulation. In particular, $\tau$ has to be small enough to resolve the oscillations in the peaks of the spreading surfaces (see Figs.~\ref{fig n6} and \ref{fig fixednode}). However, $\tau$ also has to be large enough to enable simulations that probe the long-time spreading dynamics. A large value of $k$ is desirable because it enables longer simulations, as fewer applications of the cell circuit are required. But a large value of $k$ also leads to larger Trotter errors in the weave operators.

\begin{figure}
\includegraphics[width=9.0cm]{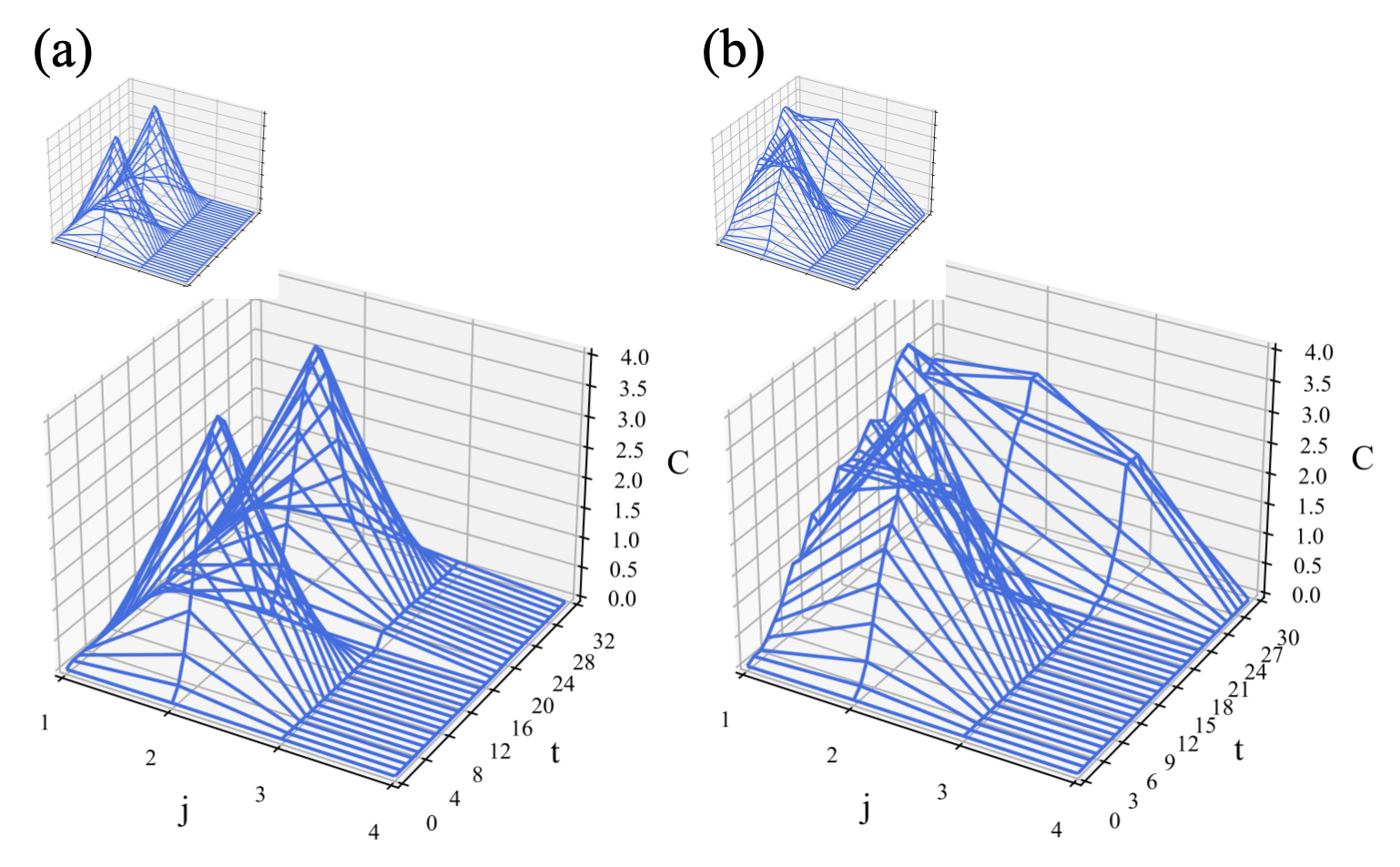} 
\caption{(Color online)  Measured operator spreading surfaces using the fixed-node OTOC and magic cell operator. (a) 16-weave in the integrable regime with $\tau = 0.098$ and up to 32 steps. (b) 20-weave in the chaotic regime with $\tau = 0.079$ and up to 30 steps. Insets show the noise-free surfaces.}
\label{fig magic}
\end{figure} 

{\it Magic cell.} In many weaving applications, the largest Trotter error will come from the cell $U(k \tau)$, because its evolution time $k \tau$ is largest. However, some simulations admit simplified Trotter steps $U(dt)$ for certain ``magic'' values of $dt$. In the Ising chain, for example, this occurs when $2 J \, dt = \pm \frac{\pi}{2},$ because for these angles we can use the decomposition
\begin{eqnarray}
R_{zz}({\textstyle \frac{\pi}{2}} )_{ij} = e^{-i\frac{\pi}{4}  Z_{i} \otimes Z_{j}}
= e^{-i\frac{\pi}{4}}  S_i S_j {\rm CZ}_{ij} , \ \ \ 
\label{magic zz rotation}
\end{eqnarray}
or its Hermitian conjugate, instead of (\ref{def zz rotation}).  Here $S \! = \!  {\rm diag}(1, e^{i \frac{\pi}{2}})$ and CZ is a controlled-$Z$ gate that can be implemented using a single CNOT and two Hadamard gates. This leads to a two-fold reduction in the number of CNOTs required. We demonstrate this variation, where the weave is built around a {\it magic} cell operator with evolution time $dt_{\rm magic} \! = k \tau = \pi / 4 |J|$. Because $dt_{\rm magic}$ is now fixed to a special value, the use of a magic cell imposes a constraint $dt_{\rm magic} \! = \! k \tau$ on $k$ and $\tau$; they are no longer independent.  Operator spreading measurements using a 16-weave with magic cell in the integrable regime, and a 20-weave with magic cell in the chaotic regime, are shown in Fig.~\ref{fig magic}. The parameter values used are shown in Table \ref{parameter table}. The data in both regimes are error mitigated. The benefits of using a magic cell here are modest, because we only apply it once. However, the use of a magic cell has the potential to significantly extend the range of Trotterized quantum simulation as gate errors improve and larger circuits become possible.

{\it Conclusions.}  Recent experiments have established that scrambling can be simulated with current gate-based quantum computers \cite{landsman2019verified,blok2021quantum,chen2020detecting,niknam2020sensitivity,mi2021information,braumuller2021probing,chen2021observation,joshi2020quantum,zhu2021observation}, making it possible to investigate interesting unsolved problems at the intersection of quantum information and physics. In this work we introduce and demonstrate techniques to enable high-resolution operator spreading measurements with cloud-based quantum computers. Trotter weaving provides the high time resolution, and the fixed-node OTOC enables larger problem sizes. Both approaches are practical for online users and have applications elsewhere in quantum information science. We observe clear signatures of operator spreading in a chaotic regime of a 4-qubit Ising model, as well as operator localization in an integrable regime. These techniques help make it possible to study information dynamics in strongly correlated and highly entangled quantum systems.

{\it Acknowledgments.} 
We thank Andreas Albrecht for useful discussions. PJC and AS acknowledge initial support and MRG acknowledges support from LANL's Laboratory Directed Research and Development (LDRD) program under project number 20190065DR. AA and ZH acknowledge, and PJC and AS acknowledge (subsequent to the above acknowledged funding) that this work was supported by the U.S. Department of Energy, Office of Science, Office of High Energy Physics QuantISED program under Contract No.  KA2401032. ZH acknowledges support from the LANL LDRD-funded Mark Kac Postdoctoral Fellowship. BY acknowledges support of the U.S. Department of Energy, Office of Science, Basic Energy Sciences, Materials Sciences and Engineering Division, Condensed Matter Theory Program, and partial support from the Center for Nonlinear Studies. 

%\bibliography{quantum.bib}

%\bibliographystyle{unsrtnat}

%\bibliography{/Users/mgeller/Dropbox/bibliographies/algorithms,/Users/mgeller/Dropbox/bibliographies/applications,/Users/mgeller/Dropbox/bibliographies/books,/Users/mgeller/Dropbox/bibliographies/cm,/Users/mgeller/Dropbox/bibliographies/dwave,/Users/mgeller/Dropbox/bibliographies/general,/Users/mgeller/Dropbox/bibliographies/group,/Users/mgeller/Dropbox/bibliographies/ions,/Users/mgeller/Dropbox/bibliographies/links,/Users/mgeller/Dropbox/bibliographies/math,/Users/mgeller/Dropbox/bibliographies/ml,/Users/mgeller/Dropbox/bibliographies/nmr,/Users/mgeller/Dropbox/bibliographies/optics,/Users/mgeller/Dropbox/bibliographies/qec,/Users/mgeller/Dropbox/bibliographies/qft,/Users/mgeller/Dropbox/bibliographies/simulation,/Users/mgeller/Dropbox/bibliographies/software,/Users/mgeller/Dropbox/bibliographies/superconductors,/Users/mgeller/Dropbox/bibliographies/surfacecode,endnotes}

%%%%%%%%%% Merge SI %%%%%%%%%%%%%%%%%%%
% Prefix an "S" to all equations, figures, tables and reset the counter 
%%%%%%%%%%%%%%%%%%%%.%%%%%%%%%%%%%%

\setcounter{equation}{0}
\setcounter{figure}{0}
\setcounter{table}{0}
\setcounter{page}{1}
\setcounter{section}{0}
\setcounter{secnumdepth}{4}
\makeatletter
\renewcommand{\thesection}{\arabic{section}}
\renewcommand{\theequation}{S\arabic{equation}}
\renewcommand{\thefigure}{S\arabic{figure}}
\renewcommand{\bibnumfmt}[1]{[S#1]}
\clearpage
\onecolumngrid
\begin{center}
\Large{ Supplementary Information for \\ ``Quantum Simulation of Operator Spreading in the Chaotic Ising Model''}
\end{center}

\vspace{1cm}
\twocolumngrid

This document provides additional details about the experimental results.  Section~\ref{Qubits} describes the online superconducting qubits used, and gives calibration results (gate errors, coherence times, and single-qubit measurement  errors) provided by the backend. Section~\ref{Raw data} presents raw data. In Sec.~\ref{Error mitigation} we discuss the error mitigation techniques used and their effect on the data. In Sec.~\ref{F0} we calculate the OTOC for the classical Ising model $H^0$.
Alternative OTOCs are discussed in Sec.~\ref{Other OTOCs}.

%This document provides additional details about the experimental results. In Sec.~\ref{Qubits} we describe the online superconducting qubits used in the experiment and give calibration results (gate errors, coherence times, and single-qubit measurement  errors) provided by the backend. In Sec.~\ref{Raw data} we give the raw data. In Sec.~\ref{Error mitigation} we discuss the error mitigation techniques used and their effect on the data. And in Sec.~\ref{F0} we calculate the OTOC for the classical Ising model $H^0$.

%

\section{Qubits}
\label{Qubits}

\begin{figure}
\includegraphics[width=8.0cm]{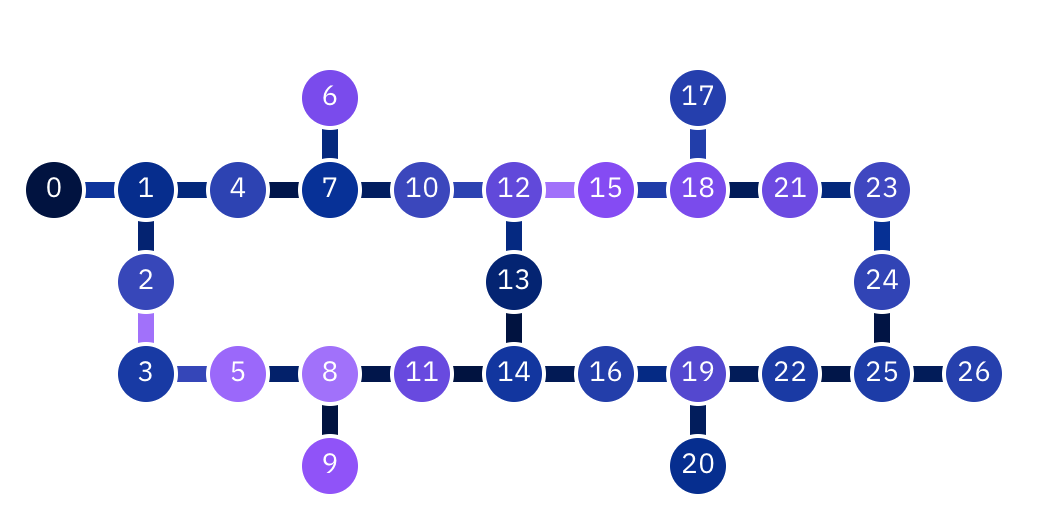} 
\caption{Layout of IBM Q device ibmq\_sydney. In this work we use qubits $Q_1$, $Q_2$, $Q_3$ and $Q_5$.}
\label{sydney chain}
\end{figure}

In this section we discuss the online superconducting qubits used in this work. Data was taken on the IBM Q processor ibmq\_sydney using the {BQP} software package developed by the authors. We measure the OTOC on qubits $\{ Q_1, Q_2, Q_3, Q_5\}$ shown in Fig.~\ref{sydney chain}. Calibration data supplied by the backend is summarized in Table \ref{calibrationDataTable}. Here $T_{1,2}$ are the standard Markovian decoherence times, and
\begin{equation}
\epsilon = \frac{T(0|1) + T(1|0)}{2}
\end{equation}
is the single-qubit state-preparation and measurement (SPAM) error, averaged over initial classical states.  $T(0|1)$ is the probability of measuring a $|0\rangle$ state after preparing $|1\rangle$; $T(1|0)$  is the reverse. The $U_2$ error column gives the single-qubit gate error measured by randomized benchmarking. The reported CNOT errors are also measured by randomized benchmarking.
 
\begin{table}[htb]
\centering
\caption{Calibration data provided by IBM Q for the ibmq\_sydney chip during the period of data acquisition.}
\begin{tabular}{c}
\begin{tabular}{|c|c|c|c|c|}
\hline
Qubit & $T_1 \ (\mu s)$ & $T_2  \ (\mu s)$ & SPAM error $\epsilon$ &  $U_2$ error \\
\hline 
$Q_1$ & 99.7  & 37.3 & 0.043 & 2.83e-4  \\
$Q_2$ & 107.9 & 65.5& 0.015 &1.80e-4  \\
$Q_3$ & 78.7 & 51.7& 0.017 & 3.15e-4  \\
$Q_5$ & 147.4 & 61.1 & 0.017 & 1.64e-4  \\
\hline 
\end{tabular}
\\
\begin{tabular}{|c|c|}
\hline 
CNOT gates &  CNOT error  \\
\hline 
\begin{tabular}{c|c} ${\rm CNOT}_{1,2}$ &  ${\rm CNOT}_{2,1}$  \\ \end{tabular} &   7.67e-3     \\
\hline 
\begin{tabular}{c|c} ${\rm CNOT}_{2,3}$ &  ${\rm CNOT}_{3,2}$  \\ \end{tabular} &   7.00e-3   \\
\hline 
\begin{tabular}{c|c} ${\rm CNOT}_{3,5}$ &  ${\rm CNOT}_{5,3}$  \\ \end{tabular} &   7.68e-3     \\
\hline 
\end{tabular} 
\\
\end{tabular}
\label{calibrationDataTable}
\end{table}

\section{Raw operator spreading data}
\label{Raw data}

\begin{figure}
\includegraphics[width=8.0cm]{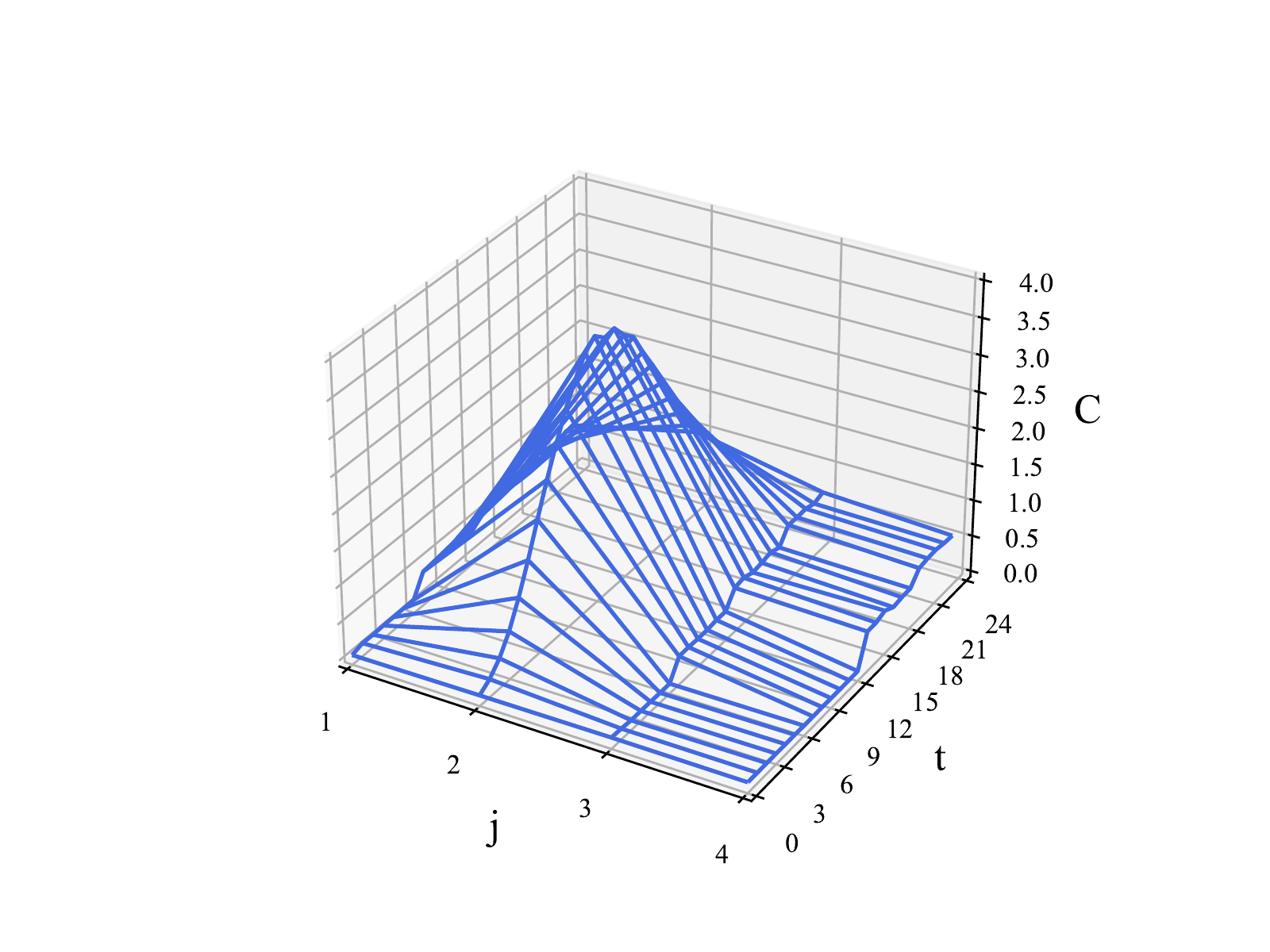} 
\caption{Raw spreading data in the integral regime, with the same settings as in Fig.~\ref{fig integrable k6 corrected}.}
\label{integrable k6 raw}
\end{figure}

\begin{figure}
\includegraphics[width=8.0cm]{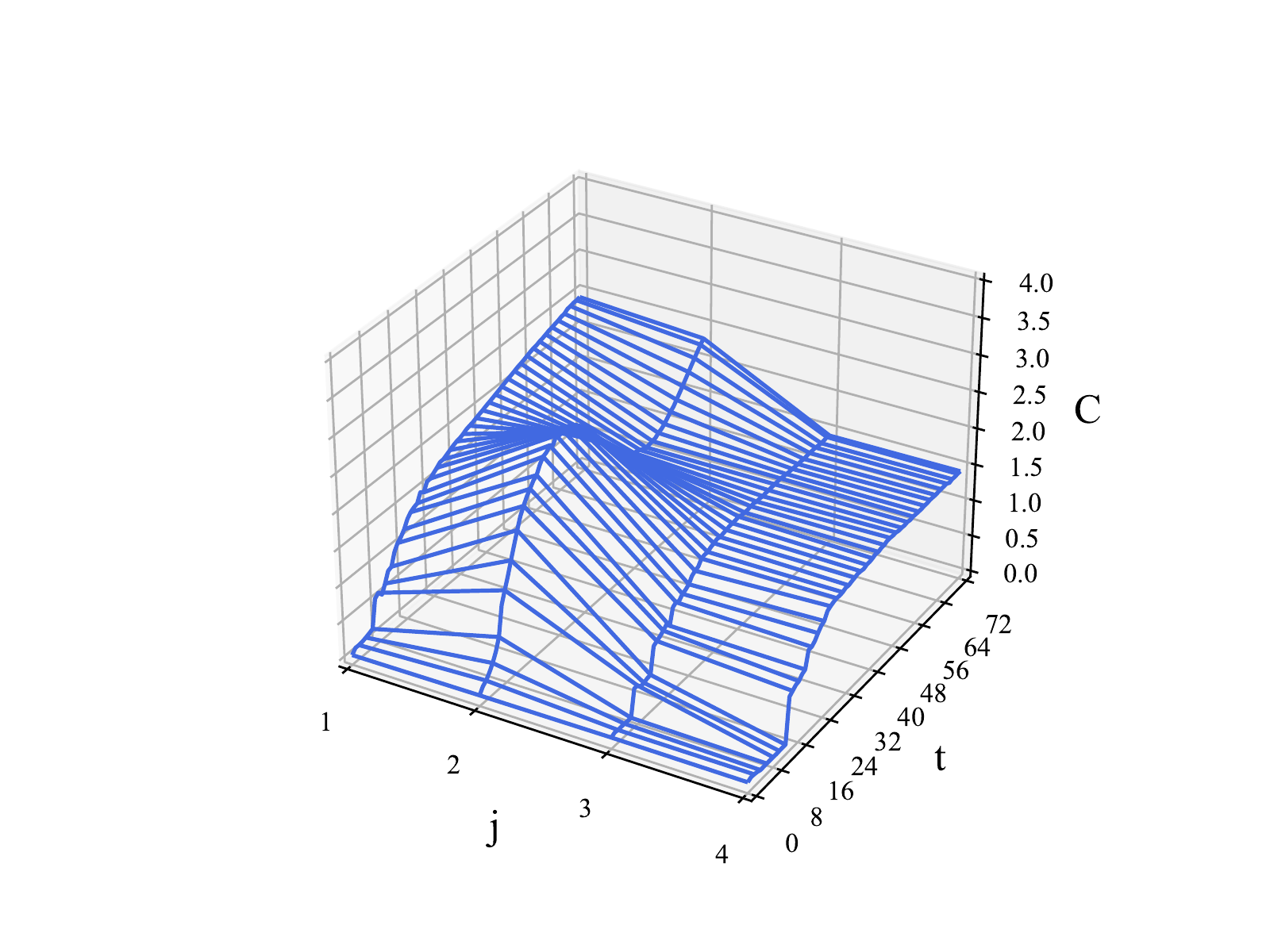} 
\caption{Raw spreading data in the chaotic regime, with the same settings as in Fig.~\ref{fig chaotic k6 corrected}.}
\label{chaotic k6 raw}
\end{figure}

The raw operator spreading surfaces are shown in Figs.~\ref{integrable k6 raw} and \ref{chaotic k6 raw}. The differences between the raw and error-mitigated data are shown in Sec.~\ref{Error mitigation}.

\clearpage

\begin{figure}
\includegraphics[width=8.0cm]{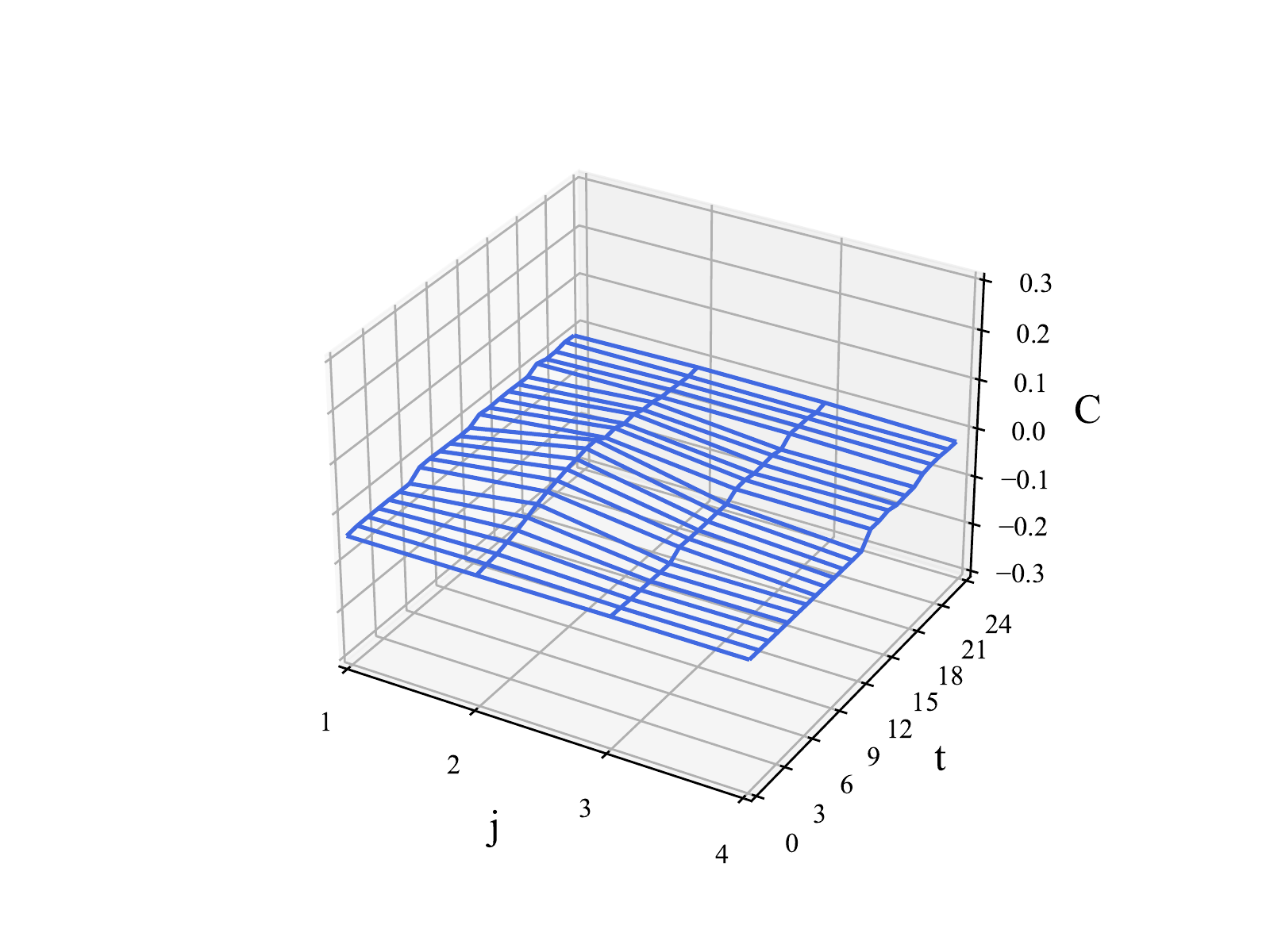} 
\caption{Difference $C_{\rm TMEM}  \! - \!  C_{\rm raw}$ between measurement-corrected and raw surfaces in the integrable regime.}
\label{integrable k6 tmatrix minus raw}
\end{figure} 

\begin{figure}
\includegraphics[width=8.0cm]{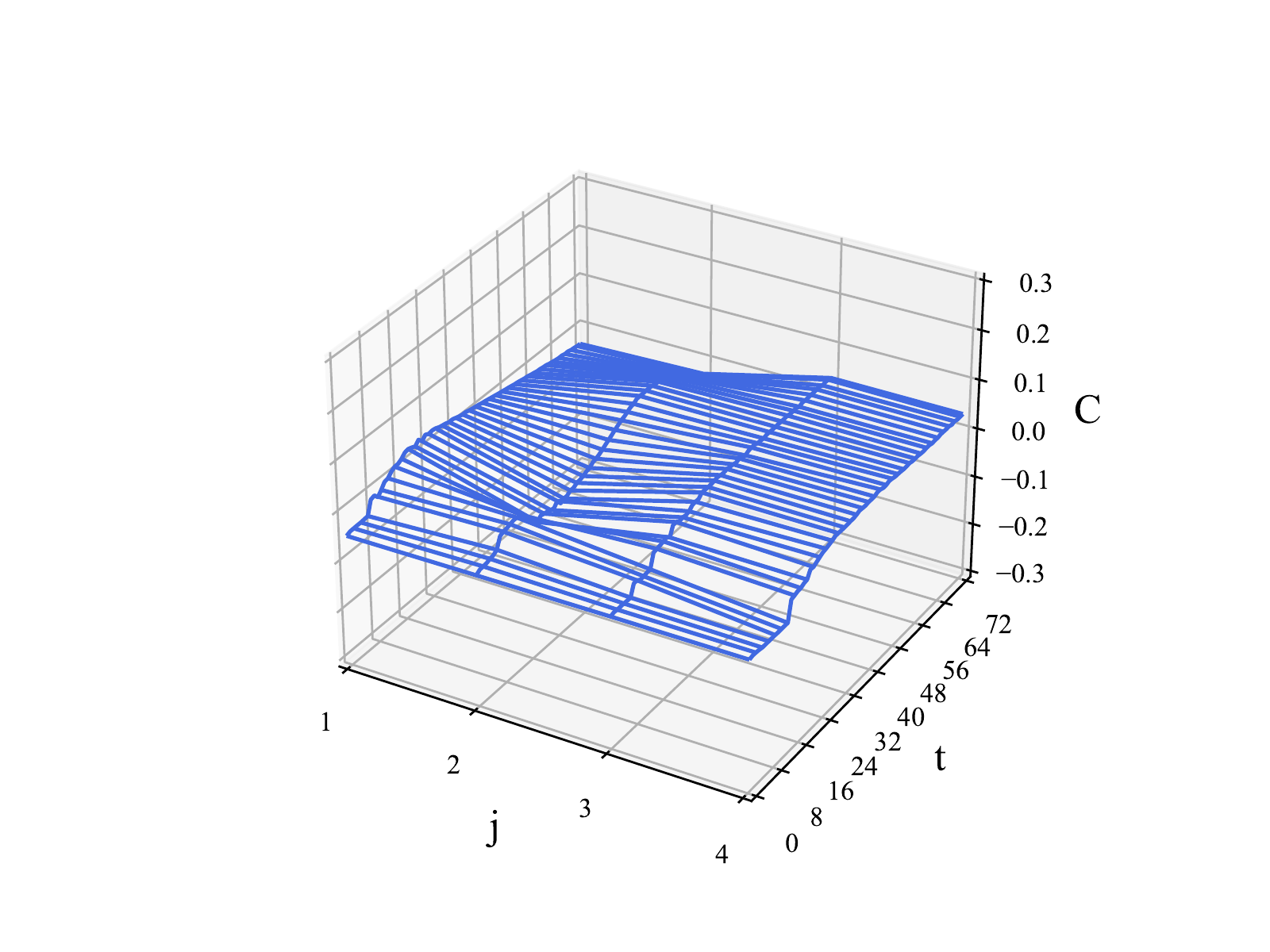} 
\caption{Difference $C_{\rm TMEM} \! - \! C_{\rm raw}$ between measurement-corrected and raw surfaces in the chaotic regime.}
\label{chaotic k6 tmatrix minus raw}
\end{figure} 

\section{Error mitigation}
\label{Error mitigation}

In this work we correct all data for measurement errors using transition matrix error mitigation (TMEM)\cite{maciejewski2020mitigation,nachman2020unfolding,hamilton2020scalable,bravyi2021mitigating,geller2021toward}, and for incoherent CNOT errors using zero-noise extrapolation \cite{temme2017error,dumitrescu2018cloud}.  In TMEM, the matrix $T$ of transition probabilities between all prepared and observed classical states $x \in \{0,1\}^n$ is initially measured. Then noisy data is corrected by minimizing  $\| T \, p_{\rm corr} - p_{\rm noisy} \|_2^2$ subject to constraints $0 \le p_{\rm corr}(x) \le 1$ and $\| p_{\rm corr} \|_1 =1$. Here $\| \cdot  \|_2$ is the Euclidean norm and $\| \cdot  \|_1$ is the $\ell_1$-norm. The effect of TMEM on the operator spreading data is shown in Figs.~\ref{integrable k6 tmatrix minus raw} and \ref{chaotic k6 tmatrix minus raw}.

\begin{figure}
\includegraphics[width=8.0cm]{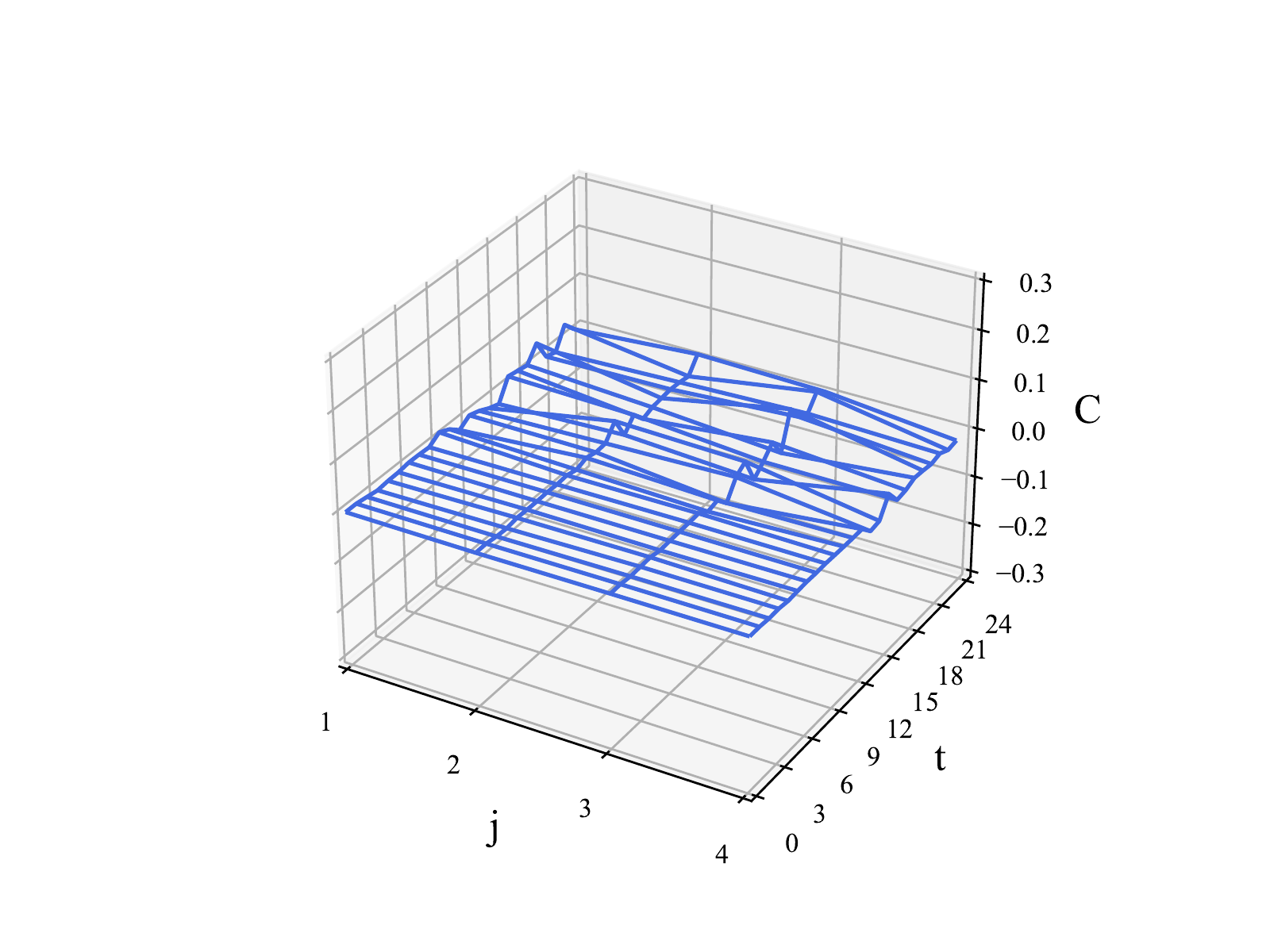} 
\caption{Difference $C_{\rm CNOT}  \! - \!  C_{\rm raw}$ between CNOT-corrected and raw surfaces in the integrable regime.}
\label{integrable k6 cnot minus raw}
\end{figure} 

\begin{figure}
\includegraphics[width=8.0cm]{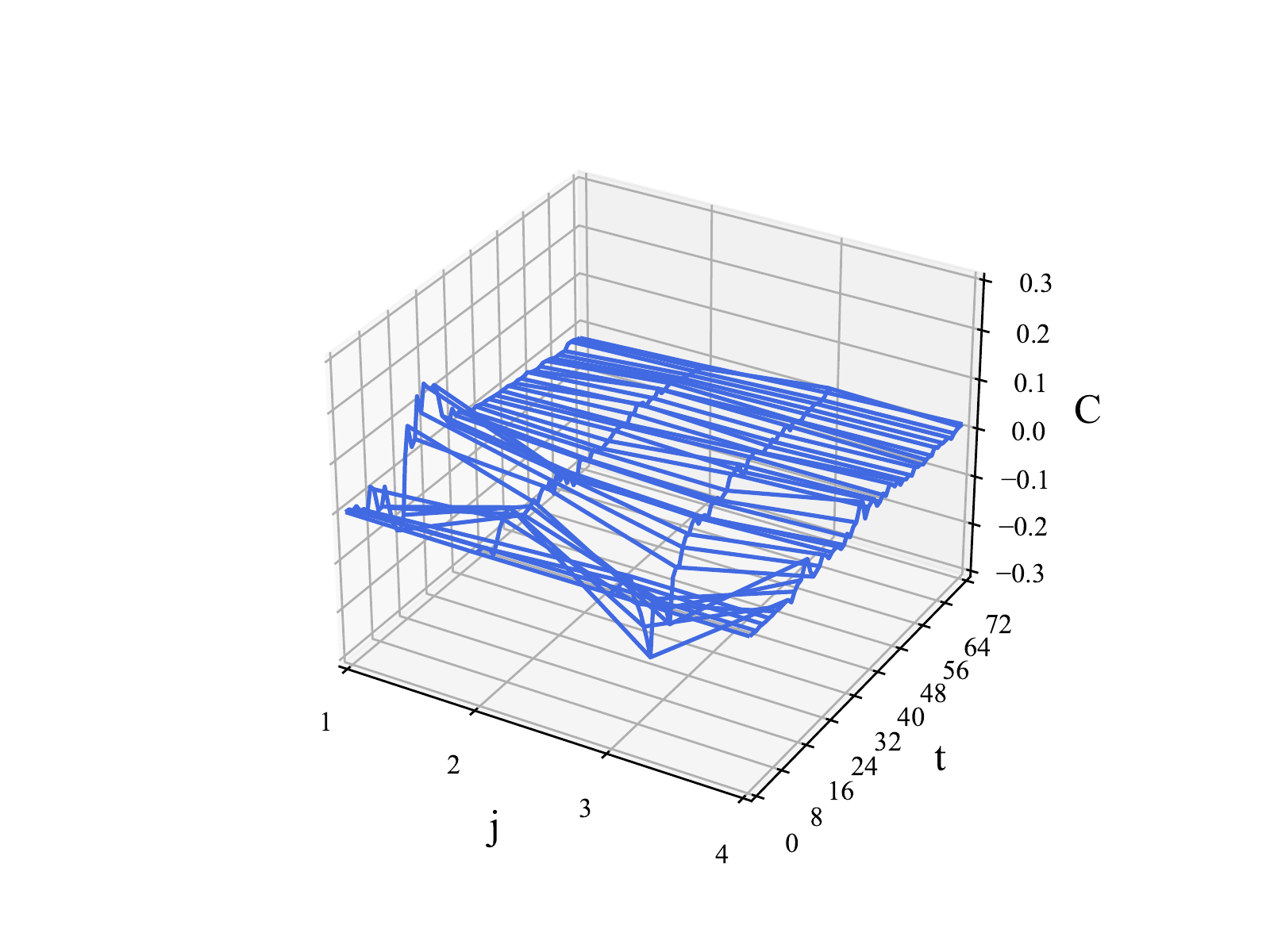} 
\caption{Difference $C_{\rm CNOT}  \! - \!  C_{\rm raw}$ between CNOT-corrected and raw surfaces in the chaotic regime.}
\label{chaotic k6 cnot minus raw}
\end{figure} 

The CNOT error mitigation involves measuring each circuit together with a variant, obtained by replacing each CNOT in the original circuit by a logically equivalent CNOT$^3$. Let ${\rm Pr}(x|m)$ be the probability of observing classical state $x$ after implementing a circuit with each original CNOT replaced by $m$ consecutive CNOTs. Then let $\{ {\rm Pr}(x|1) \}_x$ be a measured probability distribution for the original circuit, and $\{ {\rm Pr}(x|3) \}_x$ be that for the CNOT$^3$ variant. Assuming that the total incoherent CNOT error in the circuit occurs in proportion to the number of CNOTS implemented, one can use the two data points ${\rm Pr}(x|1)$  and ${\rm Pr}(x|3)$ to define a line with intercept
\begin{eqnarray}
{\rm Pr}(x|0) := \frac{ 3 \, {\rm Pr}(x|1) -   {\rm Pr}(x|3)}{2},
\end{eqnarray}
which defines a candidate correction. Here $m=0$ means no incoherent CNOT error. If $0 \le {\rm Pr}(x|0) \le 1$ for all $x$,  then we accept it as the corrected probability distribution:
\begin{eqnarray}
{\rm Pr}(x)_{\rm corr} = {\rm Pr}(x|0). 
\end{eqnarray}
Otherwise we find the physical ${\rm Pr}(x)_{\rm corr}$ closest to ${\rm Pr}(x|0)$ in Frobenius distance. The effect of CNOT noise extrapolation on the operator spreading data is shown in Figs.~\ref{integrable k6 cnot minus raw} and \ref{chaotic k6 cnot minus raw}.

\begin{figure}
\includegraphics[width=8.0cm]{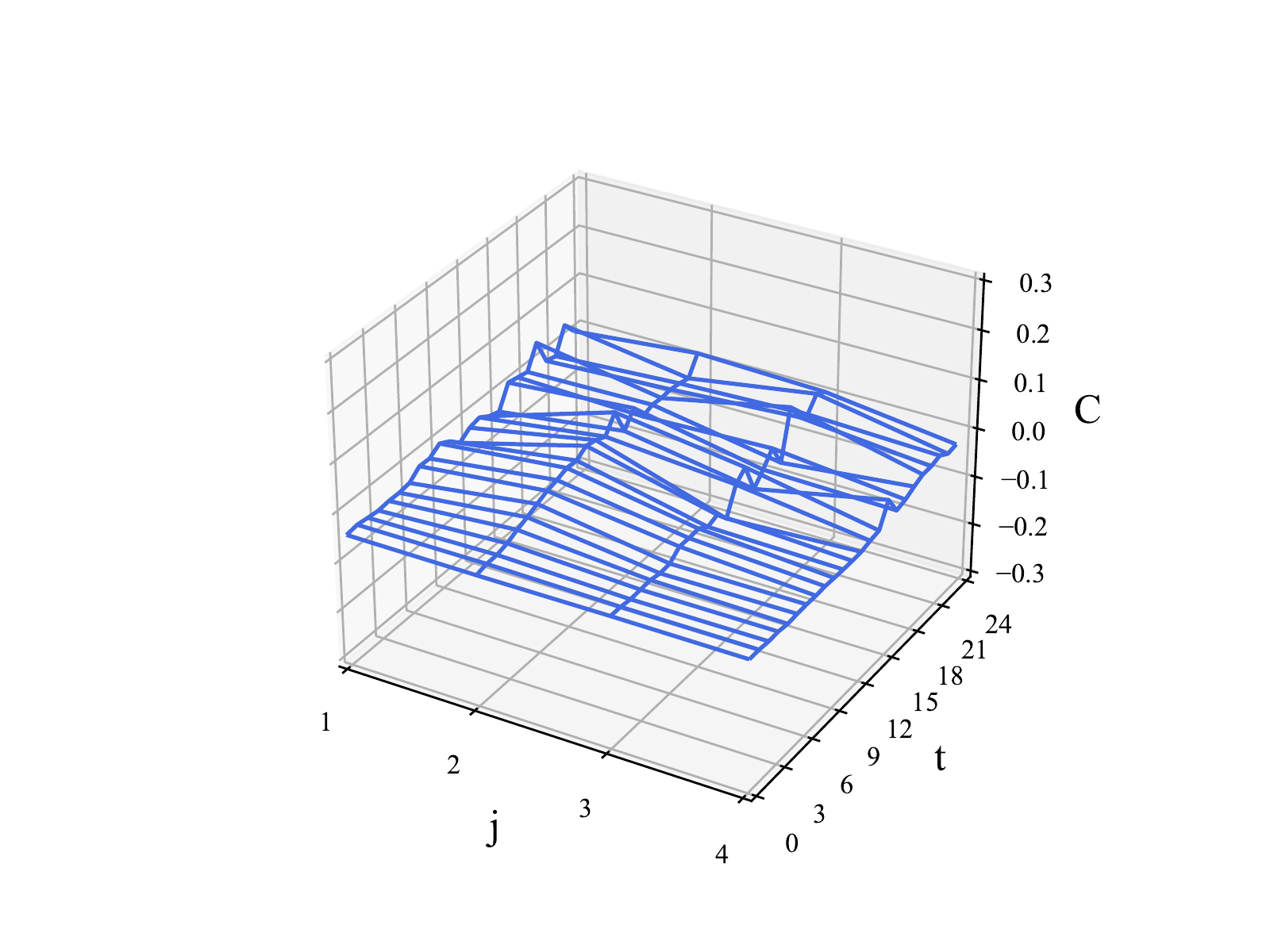} 
\caption{Difference $C_{\rm corr}  \! - \!  C_{\rm raw}$ between fully corrected and raw surfaces in the integrable regime.}
\label{integrable k6 corr minus raw}
\end{figure} 

\begin{figure}
\includegraphics[width=8.0cm]{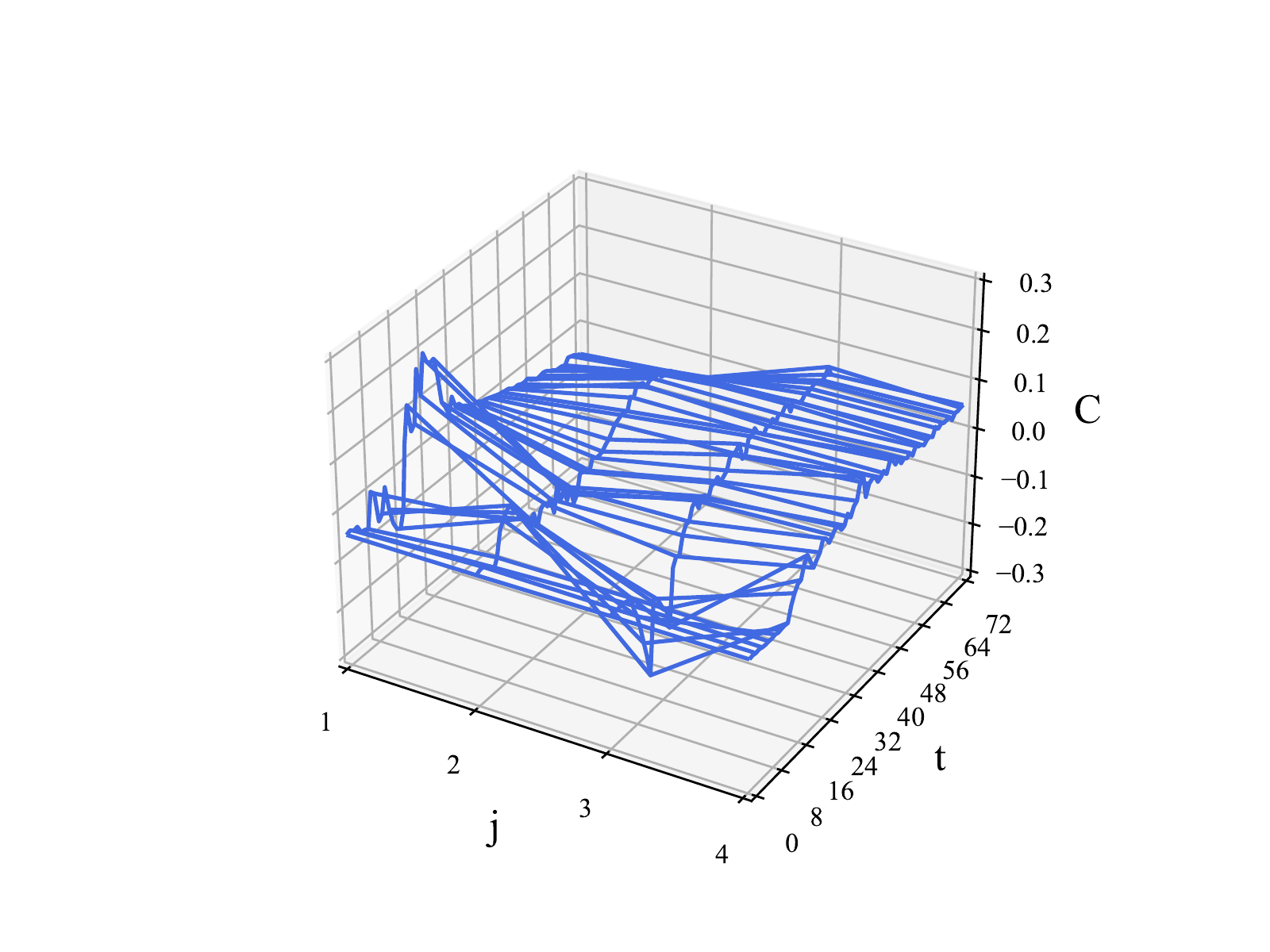} 
\caption{Difference $C_{\rm corr}  \! - \!  C_{\rm raw}$ between fully corrected and raw surfaces in the chaotic regime.}
\label{chaotic k6 corr minus raw}
\end{figure} 

The net effect of TMEM and CNOT noise extrapolation on the operator spreading data is shown in Figs.~\ref{integrable k6 corr minus raw} and \ref{chaotic k6 corr minus raw}. When both error mitigations are applied to the raw operator spreading data, we obtain the surfaces in Figs.~\ref{fig integrable k6 corrected} and \ref{fig chaotic k6 corrected}. 

\section{Classical OTOC}
\label{F0}

In the fixed-node approach, the absolute value of the OTOC is measured on the quantum processor but its phase is efficiently calculated from $F_{ij}^0$, the OTOC (\ref{def OTOC}) calculated with the classical Hamiltonian $H^0$ of (\ref{defH0}). To calculate this phase, we need the energy of the $n$-qubit state $| 0 \cdots 0\rangle$, which is
\begin{eqnarray}
E_0 = (n-1)J + n B_z.
\end{eqnarray}
Next we assume that $n \ge 3$. The energy of the single-excitation state $X_j | 0 \cdots 0\rangle = |0 \cdots 1_{j}  \cdots 0\rangle $ is
\begin{eqnarray}
E_{0 \cdots 1_{j}  \cdots 0} =
\begin{cases}
E_0 - 2 J - 2 B_z & {\rm if} \  j \in\{1,n\} , \\
E_0 - 4 J - 2 B_z & {\rm else}.  \\
\end{cases}
\label{single-excitation energies}
\end{eqnarray}
We will also need the energy of the double-excitation state $|10 \cdots 1_{j}  \cdots 0\rangle$, with $ j \ge  2$, which is
\begin{eqnarray}
E_{1 0 \cdots 1_{j}  \cdots 0} =
\begin{cases}
E_0 - 2 J - 4 B_z & {\rm if} \  j =2 , \\
E_0 - 4 J - 4 B_z & {\rm if} \  j = n , \\
E_0 - 6 J - 4 B_z  & {\rm else}.  \\
\end{cases}
\label{double-excitation energies}
\end{eqnarray}

To calculate the OTOC we first write it as
\begin{equation}
F^0_{1j}(t)= \sum_x \langle 0 \cdots 0 | X_1(t) X_j |x\rangle \langle x |  X_1(t) X_j | 0 \cdots 0 \rangle ,
\end{equation}
where $|x\rangle$ is a classical state with $x \in \{0,1\}^n$. If $j=1$ we have
\begin{eqnarray}
F^0_{11}(t) 
&=& \langle 0 \cdots 0 | X_1(t) | 10 \cdots 0 \rangle^2 \nonumber  \\
&=& \langle 0 \cdots 0 | e^{iE_0t} X_1 e^{-i(E_0 - 2J - 2B_z)t}   |10 \cdots 0 \rangle^2 \nonumber  \\
&=& e^{4i(J + B_z)t} ,
\end{eqnarray}
where we have used (\ref{single-excitation energies}). 

\begin{widetext}

\noindent If $j >1$, we have
\begin{eqnarray}
F^0_{1j}(t) &=& \sum_x \langle 0 \cdots 0 | X_1(t) X_j |x\rangle \langle x |  X_1(t) X_j | 0 \cdots 0 \rangle \\
&=& 
\langle 0 \cdots 0 | X_1(t)  | 1 0 \cdots 0 \rangle \langle 1 0 \cdots 1_j \cdots 0  |  X_1(t) | 0 \cdots 1_j \cdots 0 \rangle \\
&=& 
\langle 0 \cdots 0 | e^{i E_0 t} X_1 e^{-iHt} | 1 0 \cdots 0 \rangle \langle 1 0 \cdots 1_j \cdots 0  |  e^{iHt} X_1 e^{-iHt} | 0 \cdots 1_j \cdots 0 \rangle \\
&=& 
e^{2i(J+B_z)t}  \langle 1 0 \cdots 1_j \cdots 0  |  e^{iHt} X_1 e^{-iHt} | 00 \cdots 1_j \cdots 0 \rangle ,
\label{case i>1}
\end{eqnarray}
and there are three cases to consider: When $j =2$, (\ref{case i>1}) becomes
\begin{eqnarray}
F^0_{12}(t) &=& e^{2i(J+B_z)t}  \langle 1 1 0 \cdots 0  |  e^{iHt} X_1 e^{-iHt} | 01    \cdots 0 \rangle \\
&=& e^{2i(J+B_z)t}  \langle 1 1 0 \cdots 0  |  e^{i(E_0-2J-4B_z)t} X_1 e^{-i(E_0 - 4J - 2B_z)t} | 01    \cdots 0 \rangle \\
&=& e^{2i(J+B_z)t} e^{2i(J-B_z)t}   \\
&=& e^{4iJt},
\end{eqnarray}
where we have used (\ref{double-excitation energies}). When $2 < j < n,$ (\ref{case i>1}) becomes
\begin{eqnarray}
F^0_{1j}(t)  &=& e^{2i(J+B_z)t}  \langle 1 0 \cdots 1_j \cdots 0  |  e^{iHt} X_1 e^{-iHt} | 0 \cdots 1_j  \cdots 0 \rangle \\
&=& e^{2i(J+B_z)t}  \langle 1 0 \cdots 1_j \cdots 0  |  e^{i(E_0 - 6J - 4B_z)t} X_1 e^{-i(E_0 - 4J - 2B_z)t} | 0 \cdots 1_j \cdots 0 \rangle \\
&=& e^{2i(J+B_z)t}  e^{-2i(J+B_z)t} \\
&=& 1.
\end{eqnarray}
And when $j=n$, (\ref{case i>1}) becomes
\begin{eqnarray}
F^0_{1n}(t)  &=& e^{2i(J+B_z)t}  \langle 1 0 \cdots 01  |  e^{iHt} X_1 e^{-iHt} | 0 \cdots 01 \rangle \\
&=& e^{2i(J+B_z)t}  \langle 1 0 \cdots 01  |  e^{i(E_0 - 4J - 4B_z)t} X_1 e^{-i(E_0 - 2J - 2B_z)t} | 0 \cdots 01 \rangle \\
&=& e^{2i(J+B_z)t}  e^{-2i(J+B_z)t} \\
&=& 1.
\end{eqnarray}
Combining these results leads to the expression (\ref{F0 result}).

\end{widetext}

\section{Other OTOCs}
\label{Other OTOCs}

\begin{figure}
\includegraphics[width=8.0cm]{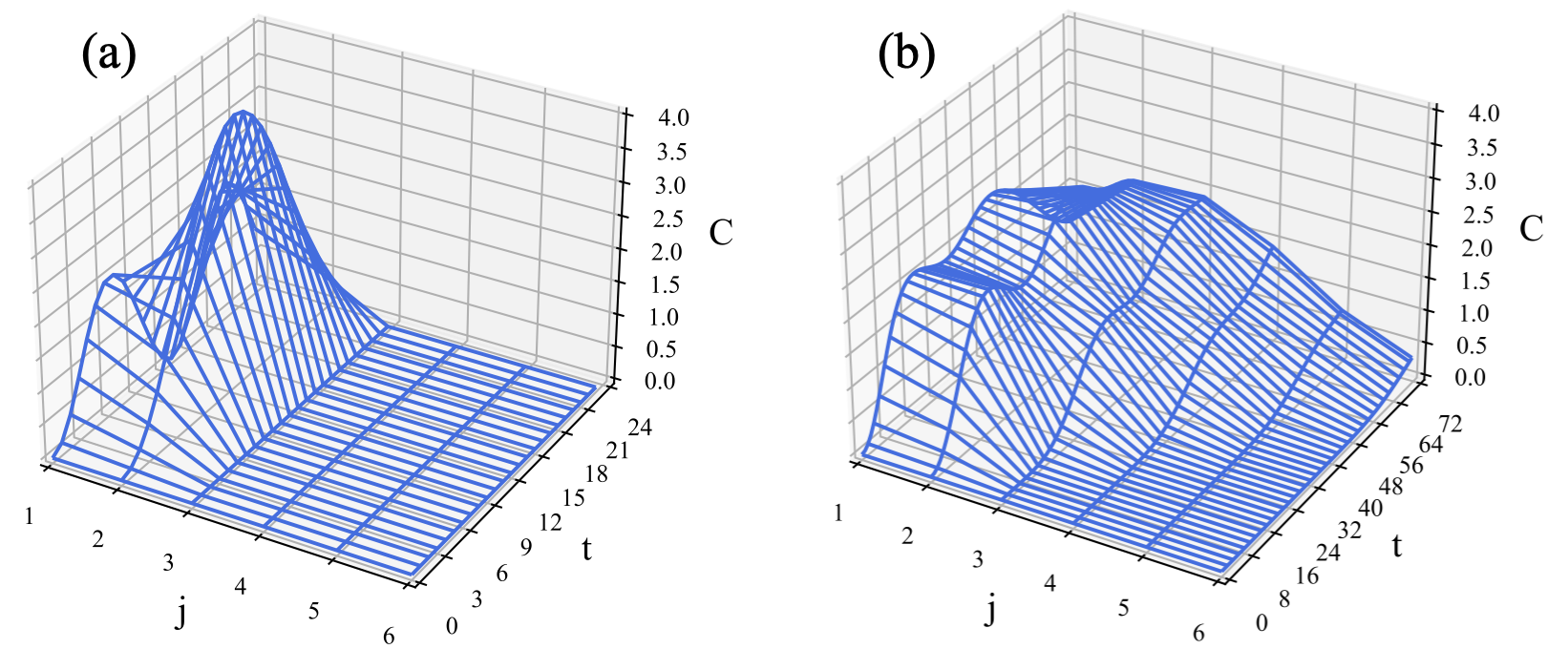} 
\caption{Infinite-temperature commutator versus qubit position $j$ and time $t$. (a) Integrable regime. The OTOC is calculated at times $ t \in \{ 0, \tau, 2 \tau , \cdots, 24 \tau \}$, with resolution $\tau \! = \!  0.06$, and time is plotted in units of  $\tau$.  (b) Chaotic regime, calculated with $\tau \! = \! 0.03$. Parameters for both regimes are given in Table \ref{parameter table}. These results were obtained by classical simulation.}
\label{n6 XX I}
\end{figure}

The operator spreading measurements in this work are based on the commutator
\begin{eqnarray}
{\rm tr}(\rho \big| [X_{i}(t), X_{j}(0)] \big|^2 ) \ \ {\rm with} \ \ \rho = |0000\rangle \langle 0000|. \ \ \
\label{def standard OTOC}
\end{eqnarray}
It is interesting to compare (\ref{def standard OTOC}) with alternative definitions. A common alternative is the infinite-temperature version
\begin{eqnarray}
{\rm tr}(\rho \big| [X_{i}(t), X_{j}(0)] \big|^2 ) \ \ {\rm with} \ \ \rho = \frac{I}{d}, \ \ 
\label{OTOC infinite temperature}
\end{eqnarray}
where $I$ is the $d \times d$ identity with $d = 2^n$. The operator spreading surfaces of Fig.~\ref{fig n6}, reevaluated with (\ref{OTOC infinite temperature}), are shown in Fig.~\ref{n6 XX I}. These surfaces are ideal results obtained classically. Overall the spreading is similar to Fig.~\ref{fig n6}, but there are detailed differences.

\begin{figure}
\includegraphics[width=8.0cm]{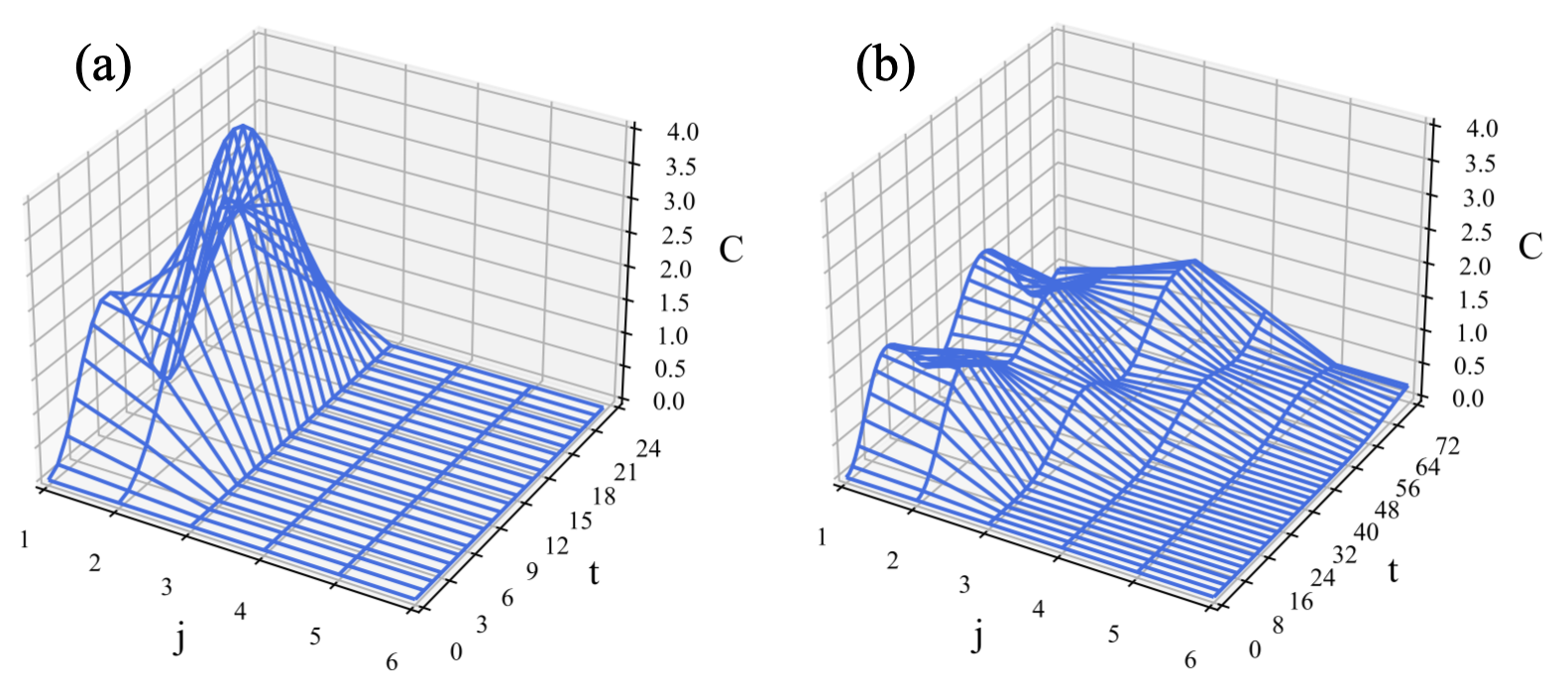} 
\caption{Operator spreading in the $|+\rangle^{\otimes n}$ state, with other parameters the same as in Fig.~\ref{n6 XX I}. }
\label{n6 XX J}
\end{figure}

It is also interesting to consider a pure state different than $|0\rangle^{\otimes n}$, such as $|+\rangle^{\otimes n}$. Figure~\ref{n6 XX J} is based on the commutator
\begin{eqnarray}
{\rm tr}(\rho \big| [X_{i}(t), X_{j}(0)] \big|^2 ) \ \ {\rm with} \ \ \rho = \frac{J}{d}, \ \ 
\label{OTOC J}
\end{eqnarray}
where $J$ is the $d \times d$ matrix of ones.

\begin{figure}
\includegraphics[width=8.0cm]{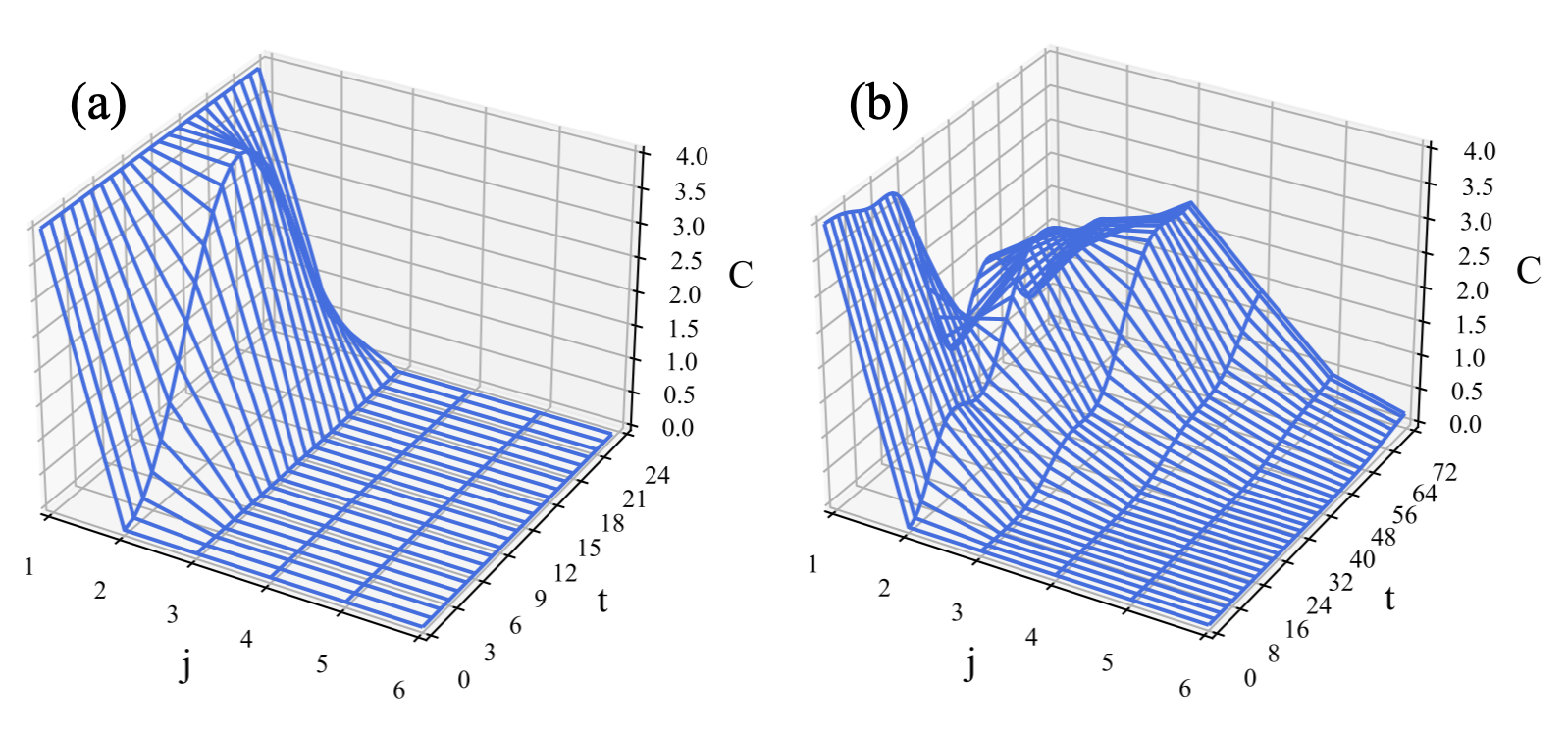} 
\caption{Spreading of $X_1$ probed by $Y_j$, with other parameters the same as in Fig.~\ref{n6 XX I}..}
\label{n6 XY}
\end{figure}

Finally, we consider the commutator
\begin{eqnarray}
{\rm tr}(\rho \big| [X_{i}(t), Y_{j}(0)] \big|^2 ) \ \ {\rm with} \ \ \rho = |0000\rangle \langle 0000|, \ \ \
\label{OTOC XY}
\end{eqnarray}
which has a peak structure different from (\ref{def standard OTOC})  because $[X_{i}(t), Y_{j}(0)] \neq 0$ at time 0. The operator spreading surfaces of Fig.~\ref{fig n6}, reevaluated with (\ref{OTOC XY}), are shown in Fig.~\ref{n6 XY}.

\end{document}